\pdfoutput=1

\documentclass[
    ,final            
  ]
  {aipproc}

\layoutstyle{6x9}

\begin{document}

\title{Search for Dark Matter from the Galactic Halo with IceCube}



\author{Carsten Rott (on behalf of the IceCube Collaboration$^{\dag}$)}{
  address={Center for Cosmology and AstroParticle Physics (CCAPP), The Ohio
    State University, Columbus, OH 43210, USA\\
    carott@mps.ohio-state.edu\\
    ${}^{\dag}$ http://www.icecube.wisc.edu}
}


\begin{abstract}
Neutrinos produced in dark matter self-annihilations in the Galactic halo
might be detectable by IceCube. We present a search for such a signal using the
IceCube detector in the 22-string configuration. We first evaluate the sensitivity before
presenting the result based on the collected data.
We find that even with the partially instrumented detector and a small
dataset, we are able to meaningfully 
constrain the dark matter self-annihilation cross-section. 
Future analyses, based on data sets from a larger detector and the
inclusion of the Galactic center, are expected to considerably improve these results. 
\end{abstract}

\maketitle



\section{Introduction}

There is overwhelming observational evidence for the existence of dark matter,
however its nature remains unknown. A variety of models predict suitable
particle candidates~\cite{Bertone:2004pz}, which typically have the properties of a weakly
interacting massive particle (WIMP).
Neutrino telescopes are powerful tools in the search for WIMPs and
their properties. They can be used to test the WIMP-nucleon scattering cross-section
similar to direct detection experiments looking for nuclear recoils.
To date, IceCube has provided very stringent constraints on the spin-dependent WIMP
nucleon scattering cross-section~\cite{Abbasi:2009uz} for WIMPs with masses
above 100~GeV by looking for neutrino signals from self-annihilating dark matter captured in the Sun.
Complementary to gamma-ray measurements, neutrino telescopes can test the dark
matter self-annihilation cross-section $\langle \sigma_{A} v \rangle$ 
(Throughout this document, we consider this product of cross-section and velocity
averaged over the dark matter velocity distribution), which is the topic of
these proceedings.

We discuss the search for dark matter
self-annihilation signals from the Galactic dark matter halo.  We first estimate the
sensitivity for the detection of such signals using the IceCube detector in
the 22-string configuration (active during 2007-2008), then present the
result on data, and comment upon future prospects.

While the result will be model independent, we have in particular sensitivity
to leptophilic dark matter in the TeV mass range, which is currently the most
compatible with the lepton excess observed by Fermi~\cite{Abdo:2009zk}, H.E.S.S.~\cite{Aharonian:2009ah}, and PAMELA~\cite{Adriani:2008zr} if interpreted
as originating from dark matter self-annihilations~\cite{Meade:2009iu}.

\section{Halo Profiles and Signal Expectations}

\begin{figure}[t]
\includegraphics[width=3.0in,height=2.5in]{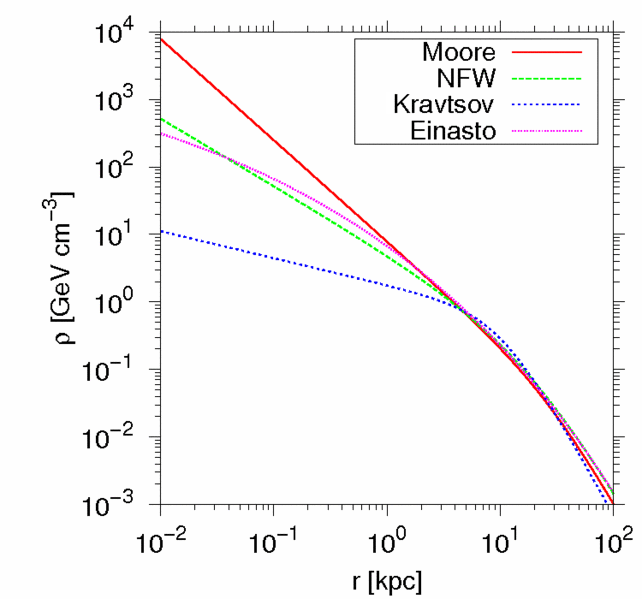}
\includegraphics[width=3.0in]{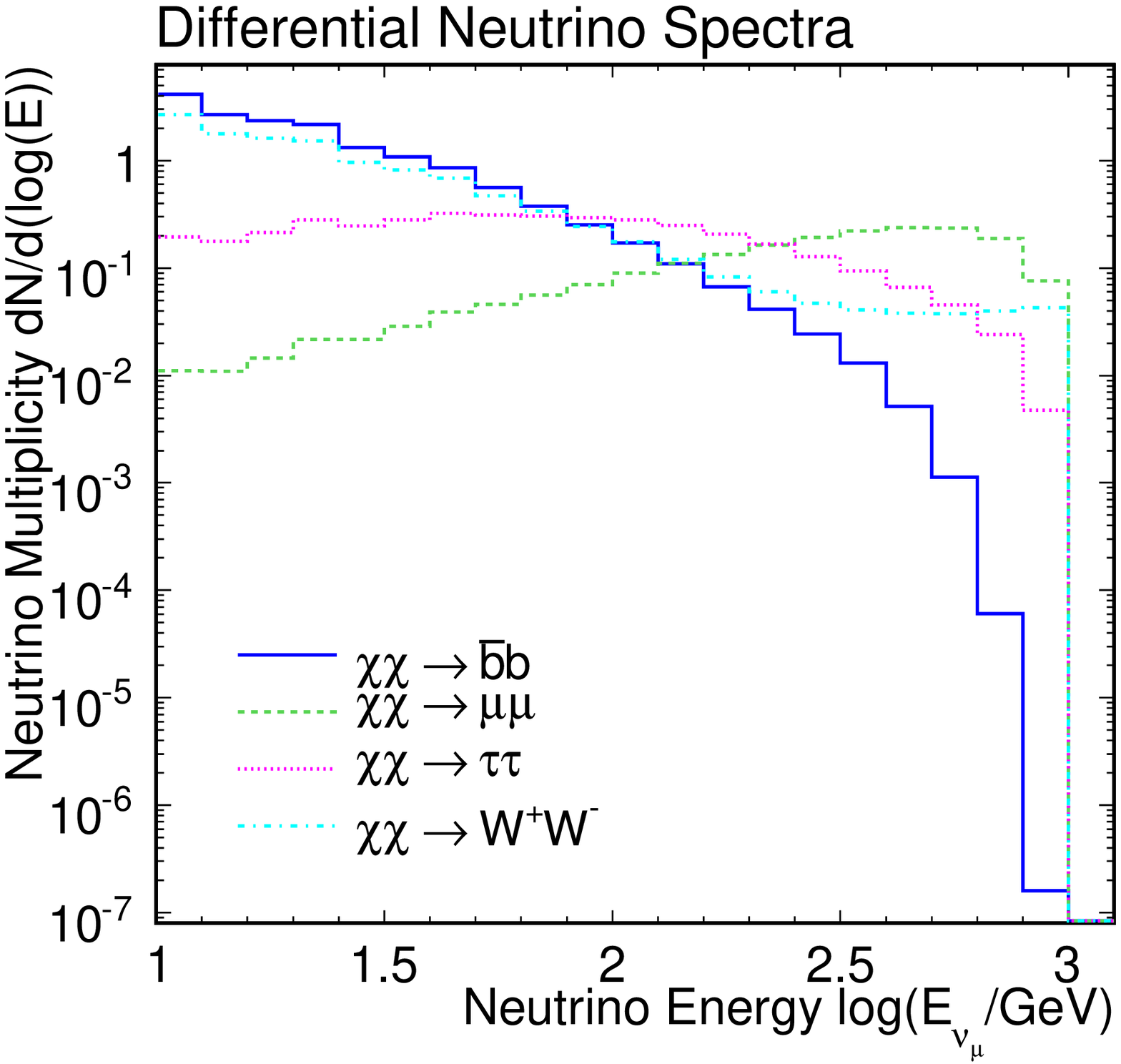}
\caption{Left: Halo dark matter density for different halo profiles. Right: Differential neutrino multiplicity per annihilation $dN_{\nu}/dE$ for
a WIMP mass of $988$~GeV.
\label{fig_halo_profiles}
\label{fig_Halo_Multi_dNdE_log}}
\end{figure}

The expected dark matter density distribution in the Milky Way can approximately be described by
spherically symmetric functions, which are motivated by fits to large scale N-body cold
dark matter simulations and observational evidence. 
Several different distribution functions and parameterizations exist. 
These halo models generally show very similar behavior for large distances from the
Galactic center (GC), however they differ significantly in their predictions close
to it. Figure~\ref{fig_halo_profiles} compares the dark matter density
profiles $\rho(r)$, as function of the distance from the Galactic center.
We use the Einasto~\cite{Einasto_profile} profile as 
benchmark model, while NFW~\cite{Navarro:1995iw}, Moore~\cite{Moore:1999gc}, and
Kravtsov~\cite{Kravtsov:1997dp} are used to estimate the uncertainty due to the halo model for this analysis.


The expected neutrino flux $\phi_{\nu}$ from dark matter self-annihilations is proportional
to the square of the dark matter density $\rho^2$ integrated along the line of sight
$J(\psi)$ for a given angular distance from the Galactic center $\psi$.
The differential neutrino flux for a WIMP of mass $m_{\chi}$ is given by~\cite{Yuksel:2007ac}:
\begin{equation}
\frac{d\phi_{\nu}}{dE} = \frac{<\sigma_A v>}{2} J(\psi) \frac{R_{sc}
  \rho_{sc}^2}{4\pi m^2_{\chi}} \frac{dN_{\nu}}{dE}.
\end{equation}
Here $R_{sc}$ and $\rho_{sc}$ are scaling factors~\cite{Yuksel:2007ac}, and $\frac{dN_{\nu}}{dE}$ is the differential
neutrino multiplicity per annihilation.

Since the annihilation products are highly model dependent, we take a 
practical approach by estimating the sensitivity for several different
annihilation channels assuming a branching ratio of 100\% for each of them in turn.
$\frac{dN_{\nu}}{dE}$ was obtained with DarkSUSY~\cite{Gondolo:2004sc} 
(see Figure~\ref{fig_Halo_Multi_dNdE_log}).
We considered the following annihilation channels:
\begin{itemize}
 \item  $\chi\chi \rightarrow b\bar{b}$ (results in a soft neutrino spectrum) 
 \item  $\chi\chi \rightarrow W^+W^-$ and $\chi\chi \rightarrow \mu\mu$
   (both result in hard neutrino spectra) 
\end{itemize}
In addition we also computed the sensitivity to a line spectrum ($\chi\chi \rightarrow \nu \nu$), which is of 
specific interest as it can be used to set a model independent limit on the
total dark matter self-annihilation cross-section~\cite{Mack:2008wu}.

\subsection{Data Selection}

We use the same
data sample and selection criteria as was used in the IceCube 22-string point
source search~\cite{Abbasi:2009iv}, but contrary to the point source analysis, which was optimized
to identify point sources with an $E^{-2}$ or $E^{-3}$ spectrum above
atmospheric neutrino background, this analysis looks for a large scale anisotropy. 
Event selection criteria have been well established, which help minimize systematic effects.
The dataset provides especially good sensitivity for annihilation signals from
high mass WIMPs.

The dataset covers the northern sky with 5114~neutrino candidate events
collected in 275.7~days of lifetime acquired during 2007-2008. It covers the 
region of $-5^\circ$ to $85^\circ$ in declination, which excludes the Galactic
center, located on the southern hemisphere at $266^\circ$~RA and $-29^\circ$~DEC.
From the Galactic center the expected neutrino flux from annihilations would be maximum as would the uncertainties due to halo models and 
other potential neutrino sources might be present (source contamination). The Galactic center is
addressed in a separate analysis using down-going starting
events~\cite{startingtracks}.

\subsection{Background Estimation and Signal Optimization}
\label{background}

We have used a simulation sample scaled to the total number of observed neutrino
candidate events to optimize the analysis. For the final data analysis
we use the data itself to estimate the background, in an effort to reduce
systematic uncertainties. We only have to rely on simulation to evaluate the signal
acceptance.


\begin{figure}[t]
\includegraphics[width=3.0in,height=2.5in]{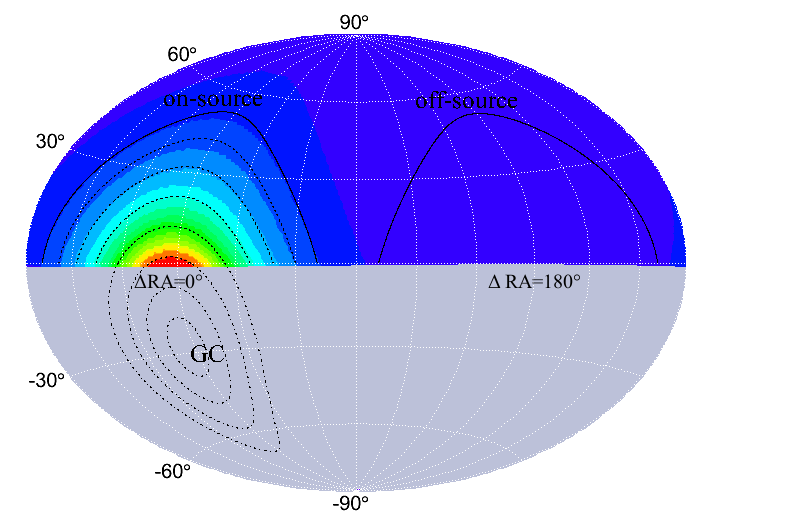}
\includegraphics[width=3.0in,height=2.5in]{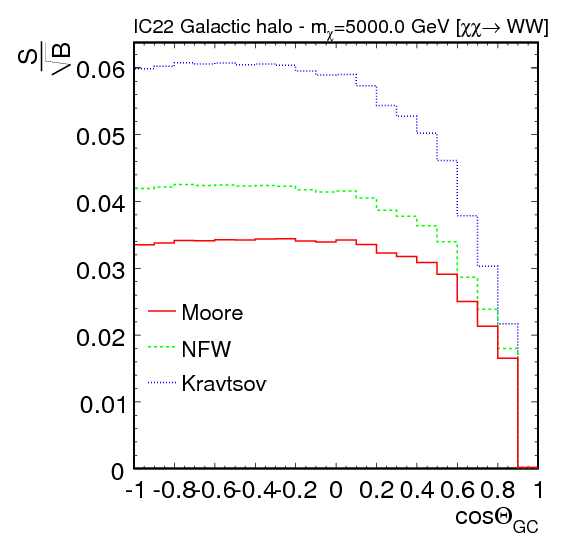}
\caption{Left: Relative expected neutrino flux from dark matter self-annihilations in the
  Milky Way halo on the northern celestial hemisphere. The largest flux is 
  expected at a RA closest to the
  Galactic center ($\Delta RA =0$). 
  Dashed lines indicate circles around the Galactic center, while the solid
  lines show the definition of on and off-source region on the northern
  hemisphere. The on-source region is centered around $\Delta RA =0$, while
  the off-source region is rotated by $180^{\circ}$ in RA. 
  Right: Optimization of the signal region as function of the distance from
  the Galactic center $\Theta_{GC}$ 
\label{fig_Skymap_North}
\label{fig_OnOffSourceRegion}
\label{fig_optimization}}
\end{figure}

A neutrino flux from dark matter self-annihilations in the Galactic dark matter halo
would manifest itself on the northern hemisphere through a large scale
anisotropy, with the largest excess neutrino flux
at the horizon and centered around the same RA as the Galactic center. 
To test this we divided the northern hemisphere in an {\it on} and {\it
  off-source} region (see Figure~\ref{fig_OnOffSourceRegion}). While the
on-source region is centered around the same RA as the GC, the off-source
region is rotated by $180^\circ$ in RA. This choice is motivated
by the robustness and simplicity of the analysis. The track reconstruction efficiency
is a function of the zenith angle, but typically averaged out in RA. We count the total
events in each region, as this makes the analysis maximally independent of
halo profiles. 
We optimized the 
size of the on-source region (as function of the angular distance from the Galactic
center $\Theta_{GC}$) to maximize $S/\sqrt(B)$ (see
Figure~\ref{fig_optimization}). Here $S$ is the expected number of signal
events from annihilations in the Galactic halo and $B$ are atmospheric neutrino
background events.
The optimal $S/\sqrt(B)$ depends on the WIMP mass and annihilation channel, but flattens out at around
$\Theta_{GC} = 80^{\circ}$, which is chosen for this
analysis. 
The maximum size of the on-source region is also constrained by the 
caveat that it should not overlap with the off-source region.
The remaining part of the observed sky (transition region), which is not classified as on or
off-source region, does not have a significant expected signal flux (compared
to the on-source region) and can be used for cross-checks.

\section{Halo Sensitivity}

The sensitivity for $\langle \sigma_A v\rangle $ was determined. 
Our atmospheric neutrino simulations predicted a Poisson distributed 1255~background
events in the equal shaped on and off-source regions.
The expected difference in events between the regions for the
'background-only' hypothesis is 
$\langle \Delta N \rangle  = \langle N_{\textit{on}}\rangle -\langle N_\textit{off}\rangle
= 0$ with an error given by $\sigma_{\Delta N} = \sqrt{2\langle
  N_{\textit{off}}\rangle }$. The expected limit (at 90\%~C.L.) on the difference
in number of events is $\Delta N_{90}=65$. 
To convert $\Delta N_{90}$ into an expected self-annihilation
cross-section limit requires accounting for the possible presence of signal in the off-source region.
\begin{equation}
\Delta N = N_{\textit{on}} - N_\textit{off} =
\left(N_{\textit{on}}(bkg)+(N_{\textit{on}}(sig)\right) -
\left(N_\textit{off}(bkg)+(N_{\textit{off}}(sig)\right)
\label{eq1}
\end{equation}
The amount of signal events in each direction depends directly on $J(\psi)$,
which would cause the difference between on and off-source regions. 

\begin{figure}[t]
\includegraphics[height=2.7in]{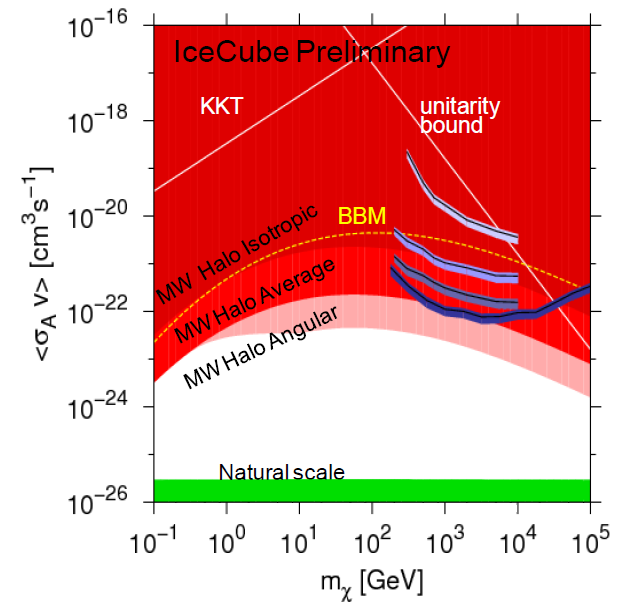}
\includegraphics[height=2.7in]{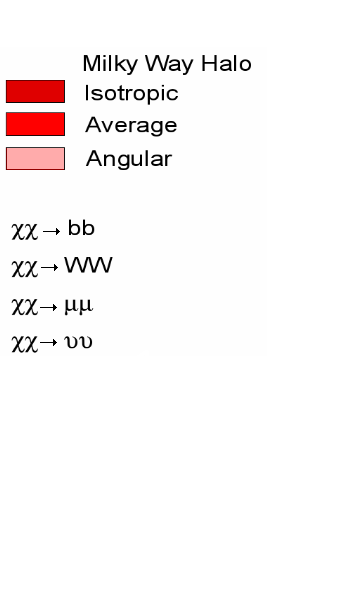}
\caption{
IceCube's sensitivity to the self-annihilation cross-section $\langle
\sigma_{A} v \rangle$ as function of the
WIMP mass $m_{\chi}$, compared to theoretical and derived limits. 
The red areas labeled ``Milky Way Halo Angular'', ``Average'', ``Isotropic'',
respectively, are  derived constraints~\cite{Yuksel:2007ac} from the observed atmospheric neutrino
energy spectrum (SuperK, Frejus, AMANDA) assuming the WIMP annihilates into
neutrinos with a 100\% branching ratio.
The green region (``Natural scale'') is for dark matter candidates consistent with being a thermal relic.
The white solid lines are the unitarity bound (right) and the cross-section
for which annihilation flattens the cusps of the halo (left).
The yellow dashed line is an estimate for a limit from cosmic neutrino
observations~\cite{Mack:2008wu}.
The blue broad lines are the IceCube expected average upper limits for 22-string detector data for
different annihilation channels. The central black line is for the Einasto
profile (NFW is almost identical), while the 
width is given by the results for the Moore and Kravtsov profile.
\label{fig_Halo_Sensitivity}}
\end{figure}

The signal expectation in both regions scales with $\langle \sigma_A v\rangle $.
The difference in the expected number of neutrino events between the on and
off source region is given by $\Delta N_{sig} = N_{\textit{on}}(sig) -
N_{\textit{off}}(sig)$. 
For a given cross-section $\langle \sigma_{A} v\rangle _0$ it is determined from simulations.
For all other cross-sections the expected number is then given by:
\begin{equation}
\Delta N_{sig}(\langle \sigma_{A} v\rangle ) = \frac{\langle \sigma_{A} v\rangle }{\langle \sigma_{A} v\rangle _{0}} (N_{on,sig}(\langle \sigma_{A} v\rangle _0) -
 N_{off,sig}(\langle \sigma_{A} v\rangle _{0})).
\end{equation}
Therefore the cross-section sensitivity at 90\% C.L. is given by $\langle \sigma_{A}
v\rangle _{90} = \Delta N_{90} \times \frac{\langle \sigma_{A} v\rangle}{\Delta N_{sig}(\langle \sigma_{A} v\rangle )}$.
Under the background only assumption we obtain IceCube's sensitivity 
for $\langle \sigma_{A} v\rangle$ as shown in Figure~\ref{fig_Halo_Sensitivity} 
as function of the WIMP mass and annihilation channel.


\section{Systematic Uncertainties}

We have performed a preliminary study of the systematic uncertainties
associated with the background estimation and signal acceptance.
By design, the comparison of events in the on and off-source regions enables 
the determination of the background from the data itself, which allows for most
systematic effects associated with understanding the detector to 
cancel out. The remaining
systematic uncertainties on the background are due to a possible large scale anisotropy of the
measured data. This could be caused by variations in the exposure for
different right ascensions, or by an existing anisotropy in the cosmic ray
flux, which translates into the atmospheric muon neutrino flux.
These two effects are expected to dominate the systematic
uncertainty on the background estimation.

The TIBET air shower array has observed an anisotropy in the cosmic ray flux in
the northern hemisphere~\cite{Amenomori:2006bx}, which is also present on the southern
hemisphere, as observed by IceCube in the down-going muon flux~\cite{IceCubeCCAPPAbbasi}. 
For cosmic ray energies of about 50~TeV, which are most relevant 
in contributing to the background atmospheric muon neutrino flux 
of this analysis, the scale of the overall anisotropy is about $0.2\%$.
In the worst case if the anisotropy is aligned so that it peaks in one
region and is minimal in the other one, this effect 
could contribute a difference of three events between on and off-source regions.
The effect on the sensitivity is estimated to be less than 4\%.

The track reconstruction efficiency varies as function of the azimuth
angle. This variation in detector coordinates\footnote{The
  track reconstruction efficiency variation is present in the partially instrumented
  22-string detector. It becomes more uniform for the full IceCube detector.} is typically averaged
out in RA, however the usage of a geo-synchronous satellite for communication
with the South Pole introduces a slight bias in the sidereal time when
maintenance runs are performed. The choice of symmetric on and off-source
regions rotated by $180^{\circ}$ in RA, 
reduces this effect significantly as 
the track reconstruction efficiency is almost identical to the case where the detector is rotated
by $\pi$.
The total expected variation in events is comparable to the cosmic ray
anisotropy effect. A correction for this effect is possible, however was not
used as one would still end up with a comparable uncertainty on an azimuth
angle dependent scale factor.

The systematic uncertainty on the signal acceptance is dominated by the DOM 
efficiency, ice properties, and a discrepancy between simulations and data 
for nearly horizontal events, which was observed in the point source analysis.
This discrepancy could be related to the DOM efficiency and ice properties
and is under further investigation.

\section{Result on Data}

After the described signal optimization and sensitivity study on simulation, we have
performed the analysis on data collected during 2007-2008 with the IceCube 22-string
configuration. This was done in a blind manner, meaning that
we only looked at the data once the selection criteria
and analysis procedure were finalized.

\begin{figure}[ht]
\includegraphics[width=3.0in]{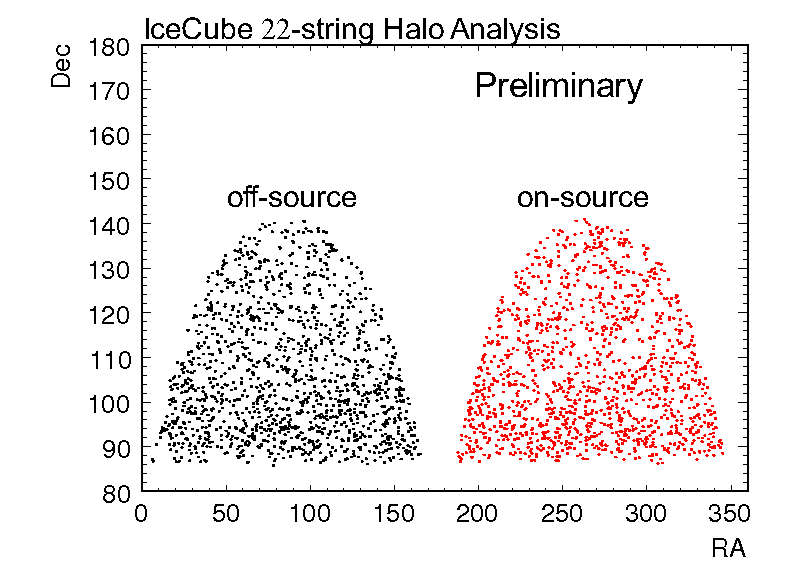}
\includegraphics[width=3.0in]{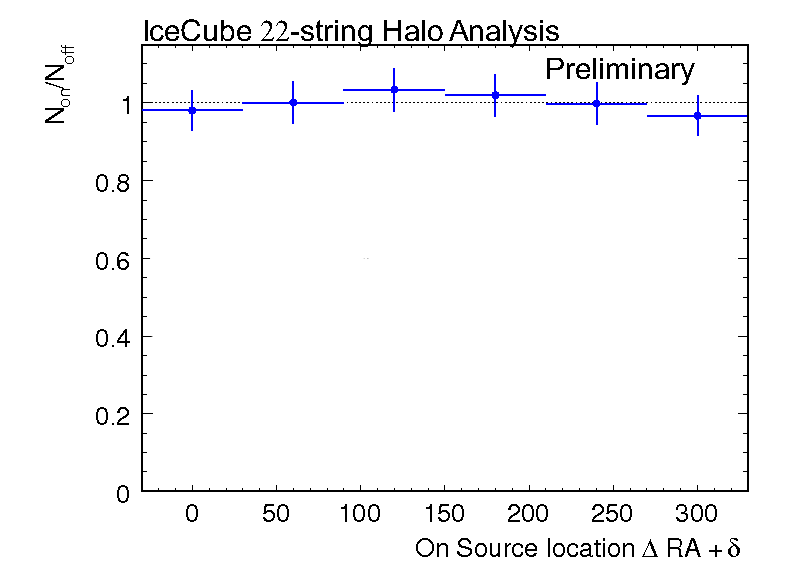}
\caption{The distribution of the neutrino candidate
events in the on (red) and off (black) source region.
We have rotated the on-source region in $60^{\circ}$ steps to
be centered at $\Delta RA + \delta$, to check that there is no systematic
effect present. Note that bins 4-6 are the inverse of bins 1-3.
\label{fig_on_off_source}}
\end{figure}

\begin{figure*}[t]
\includegraphics[width=3.0in]{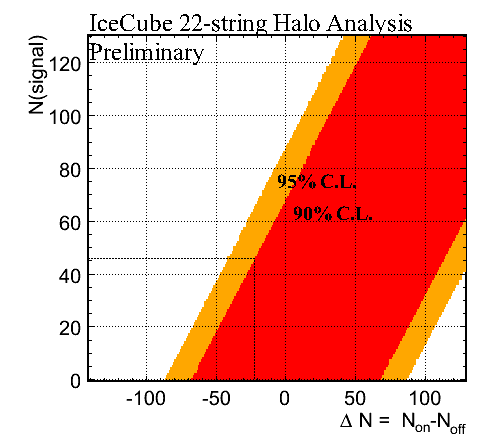}
\includegraphics[width=3.0in]{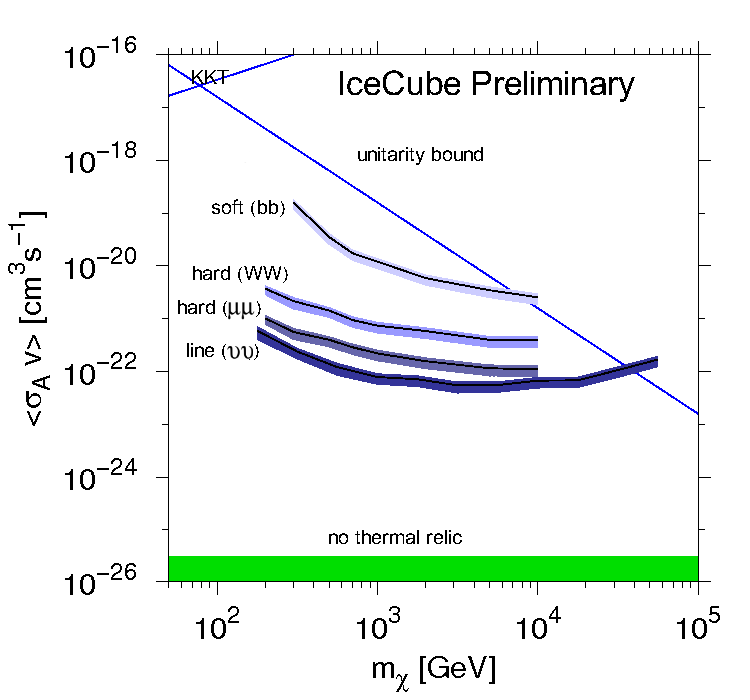}
\caption{Left:
  Confidence belt for a background expectation of 1389~events. Based on an observation of 1367~events $\Delta N_{90}$
  is determined to be 46.
  Right: Preliminary 90\% C.L. exclusion limit (no systematics included). Shown are theoretical
  bounds and the bound derived from this analysis for a given
  annihilation channel: soft, hard, line  (in order top to bottom). The central black line is for the Einasto
  profile, while the width is given by the results for the Moore and Kravtsov profile.
\label{fig_exclusion_limit}}
\end{figure*}

In data we observed 1367~events in the on-source region, while the off-source
region contains $N_{off}=1389$.
Figure~\ref{fig_on_off_source} shows the distribution of these neutrino candidate
events.
The number of observed events are consistent with each other and we have
computed a 90\% C.L. limit, by constructing the confidence interval through Monte Carlo.
The number of observed events in the off-source region is our best guess on
the background. We scan a wide range of possible outcomes and calculate the intervals 
that would contain the signal with 90\% C.L. 
Figure~\ref{fig_exclusion_limit} shows the confidence interval and the preliminary exclusion limit
obtained.

To check for any possible systematic effect in the analysis, we have rotated
the on and off-source region in $60^{\circ}$ steps.
For all the bins, no effect was observed, i.e.  the ratio of $N_{on}/N_{off}$ is consistent with one (see Figure~\ref{fig_on_off_source}).


\section{Conclusions and Discussions}

We have evaluated IceCube's sensitivity towards the detection of neutrinos
from dark matter self-annihilations in the Galactic dark matter halo by
searching for a large scale anisotropy on the northern hemisphere.
Using data collected during 2007-2008 with IceCube in the 22-string
configuration, we have not observed any such anisotropy.
With the performed analysis we were able to place relevant constraints on the 
dark matter self-annihilation cross-section $\langle \sigma_{A} v \rangle$ 
at 90\% C.L. for WIMPs in the mass the range from a few hundred
GeV to several TeV.
Such an analysis has previously not been performed with AMANDA or IceCube and $\langle \sigma_{A} v \rangle$
has only been constrained through theory or theoretical derived limits.

IceCube's reach can be significantly
improved by looking at the Galactic center, which will be possible beginning
with the IceCube 40-string dataset, by using neutrinos interacting inside the
detector volume. While such an analysis will be able to put
significantly better constraints, a large scale anisotropy would provide a more distinct
discovery signal.

%
\begin{theacknowledgments}

We thank John Beacom, Shunsaku Horiuchi, and Matt Kistler for valuable discussions.
We acknowledge the support from the following agencies: U.S. National Science
Foundation-Office of Polar Program, U.S. National Science Foundation-Physics
Division, University of Wisconsin Alumni Research Foundation, U.S. Department
of Energy, and National Energy Research Scientific Computing Center, the
Louisiana Optical Network Initiative (LONI) grid computing resources; Swedish
Research Council, Swedish Polar Research Secretariat, and Knut and Alice
Wallenberg Foundation, Sweden; German Ministry for Education and Research
(BMBF), Deutsche Forschungsgemeinschaft (DFG), Research Department of Plasmas
with Complex Interactions (Bochum), Germany; Fund for Scientific Research
(FNRS-FWO), FWO Odysseus programme, Flanders Institute to encourage scientific
and technological research in industry (IWT), Belgian Federal Science Policy
Office (Belspo); Marsden Fund, New Zealand.

\end{theacknowledgments}
%


\bibliographystyle{aipproc}   

\bibliography{Rott_HEP_Talk_bibtex}

\IfFileExists{\jobname.bbl}{}
 {\typeout{}
  \typeout{******************************************}
  \typeout{** Please run "bibtex \jobname" to optain}
  \typeout{** the bibliography and then re-run LaTeX}
  \typeout{** twice to fix the references!}
  \typeout{******************************************}
  \typeout{}
 }

\end{document}